# Vignetting effect in Fourier ptychographic microscopy

An Pan,[1,2] Chao Zuo,[3,*] Yuege Xie,[4] Yan Zhang,[1,2] Ming Lei,[1] Baoli Yao[1,*]

[1] *State Key Laboratory of Transient Optics and Photonics, Xi'an Institute of Optics and Precision Mechanics, Chinese Academy of Sciences, Xi'an 710119, China*
[2] *University of Chinese Academy of Sciences, Beijing 100049, China*
[3] *Smart Computational Imaging (SCI) Laboratory, Nanjing University of Science and Technology, Nanjing 210094, China*
[4] *Institute for Computational Engineering and Sciences (ICES), University of Texas at Austin, Austin, TX 78712, USA*
*Corresponding author: surpasszuo@163.com (C. Zuo), yaobl@opt.ac.cn (B. Yao)*

**Fourier ptychographic microscopy (FPM) is a computational imaging technique that overcomes the physical space-bandwidth product (SBP) limit of a conventional microscope by applying angular diversity illuminations. In the usual model of FPM, the microscopic system is approximated by being taken as space-invariant with transfer function determined by a complex pupil function of the objective. However, in real experimental conditions, several unexpected "semi-bright and semi-dark" images with strong vignetting effect can be easily observed when the sample is illuminated by the LED within the "transition zone" between bright field and dark field. These imperfect images, apparently, are not coincident with the space-invariant model and could deteriorate the reconstruction quality severely. In this Letter, we examine the impact of this space-invariant approximation on FPM image formation based on ray-based and rigorous wave optics-based analysis. Our analysis shows that for a practical FPM microscope with a low power objective and a large field of view, the space invariance is destroyed by diffraction at other stops associated with different lens elements to a large extent. A modified version of the space-variant model is derived and discussed. Two simple countermeasures are also presented and experimentally verified to bypass or partially alleviate the vignetting-induced reconstruction artifacts. © 2017 Optical Society of America**

Fourier ptychographic microscopy (FPM) [1-6] is a fast-growing computational imaging technique with high resolution (HR) and wide field-of-view (FOV), which shares its root with conventional ptychography [7, 8] and synthetic aperture imaging [9, 10]. Due to its flexible setup, perfect performance without mechanical scanning and interferometric measurements, FPM has wide applications in the digital pathology [11] and whole slide imaging systems [12]. Generally, the coherent microscopic system is simplified to a linear space-invariant (LSI) 4-*f* imaging system with transfer function determined by a complex pupil function of the objective and a simple convolution operation describing the object-image relationship [13, 14]. In real experimental conditions, however, the resolution may decrease from the center to the edge of FOV [15]. Several unexpected "semi-bright and semi-dark" imperfect images with strong vignetting effect can be easily observed when the sample is illuminated by the LED within the "transition zone" between bright field (BF) and dark field (DF). These imperfect images, apparently, are not coincident with the LSI model and could deteriorate the reconstruction quality of FPM severely. Especially the edge of FOV, there will be obvious wrinkles. In this letter, we examine the impact of this space-invariant approximation on FPM image formation based on ray-based and rigorous wave optics-based analysis. Our analysis shows that for a practical FPM microscope with a low power objective and a large field of view, the LSI model is destroyed by diffraction at other stops associated with different lens elements to a large extent. A modified version of the linear space-variant (LSV) model is derived and discussed. Two simple countermeasures are also presented and experimentally verified to bypass or partially alleviate the vignetting-induced reconstruction artifacts.

In the conventional analysis of 4-*f* imaging systems, it is generally assumed that even though an imaging system may consist of several optical elements, respectively with its own aperture, these elements are often lumped together in a single "black box," and only the exit or the entrance pupil is used to describe the effects of diffraction [14, 16]. In the paraxial regime, this simplified analysis indeed results in a LSI model. However, if the objective $L_1$ is not infinite in extent as shown in Fig.1 (a), the wave leaving $L_1$ is subject to the effects of diffraction by the lens aperture itself. Often the diameter of tube lens $L_2$ is greater than objective $L_1$, $L_2$ can be considered to be effectively infinite, and vignetting can be introduced by the finite aperture of the objective $L_1$ [13]. Point source $PS_{VL}$ (pink light) marks the onset of vignetting in the geometrical optics regime, and we use the subscript VL to refer to this point as the vignetting limit. For even farther off-axis point, for example, the green light in Fig.1 (a), the aperture delimiting $L_1$ eliminates part of the light and the aperture in Fourier plane is thus no longer fully illuminated. With vignetting, the region of the Fourier plane aperture that is no longer illuminated cannot contribute to the distribution in the image plane. The consequences of this diffraction depend on the location of the point source, and thus the imaging operation is space variant and must be described not by a convolution integral but rather by a more general superposition integral.

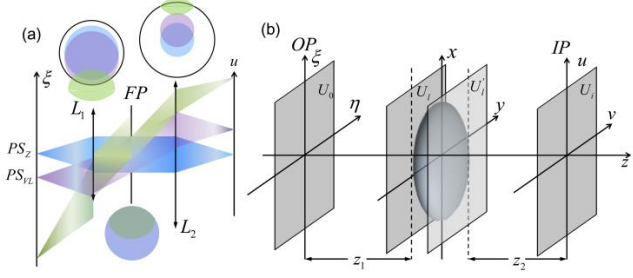

Fig. 1. (a) Example of vignetting in a simplified 4-$f$ imaging system for off-axis point (pink light) and even farther off-axis point (green light) from the ray-based analysis. FP, Fourier plane; $L_1$, the objective; $L_2$, tube lens; PS, point source. (b) A single lens system for illustration from the wave optics-based analysis. OP, object plane; IP, image plane.

Figure 1 (b) shows a single lens imaging system for illustration from the perspective of wave optics in detail. Since the wave propagation is linear, the field $U_i$ can always be represented as superposition integral as follows [14].

$$U_i(u,v) = \iint h(u,v;\xi,\eta) U_0(\xi,\eta) d\xi d\eta \quad (1)$$

where $h$ is the impulse response function (IRF) of this system. In order to get the IRF, let the object function be a $\delta$ function at the $(\xi, \eta)$ plane. Then the input and output of lens transmission function are given by

$$U_l(x,y) = \frac{1}{j\lambda z_1} \exp\left\{ j\frac{k}{2z_1}\left[(x-\xi)^2 + (y-\eta)^2\right] \right\}$$
$$U_l'(x,y) = U_l(x,y) P(x,y) \exp\left[-j\frac{k}{2f}(x^2+y^2)\right] \quad (2)$$

where $P(x, y)$ is the pupil function [14]. Then with the Fresnel transform, the optical field of image plane can be given by

$$h(u,v;\xi,\eta) = \frac{1}{j\lambda z_2}\iint U_l'(x,y)\exp\left\{ j\frac{k}{2z_2}\left[(u-x)^2+(v-y)^2\right]\right\}dxdy$$
$$= \frac{1}{\lambda^2 z_1 z_2}\exp\left[j\frac{k}{2z_2}(u^2+v^2)\right]\exp\left[j\frac{k}{2z_1}(\xi^2+\eta^2)\right]$$
$$\times \iint P(x,y)\exp\left[j\frac{k}{2}\left(\frac{1}{z_1}+\frac{1}{z_2}-\frac{1}{f}\right)(x^2+y^2)\right]$$
$$\times \exp\left\{-jk\left[\left(\frac{\xi}{z_1}+\frac{u}{z_2}\right)x+\left(\frac{\eta}{z_1}+\frac{v}{z_2}\right)y\right]\right\}dxdy \quad (3)$$

There are three quadratic phase factors. The term in integral can be easily eliminated if the equation $1/z_1+1/z_2=1/f$ holds, which is the precondition for imaging. The other two terms are outside the integral. The one which is associated with the image plane, can be ignored if we only consider the intensity. The residual term, which is associated with the object plane, is really intractable. Goodman et al. [17] have proposed two ways to compensate or ignore it, one is to illuminate with the convergent spherical wave and the other one is to use the condition that the size of object needs to be smaller than the lens less than 1/4. Thus the general IRF is well known as follows, which is the Fourier transform of pupil function.

$$h_p(u,v;\xi,\eta) \approx \iint P(x,y)\exp\left\{-j\frac{2\pi}{\lambda z_2}\left[\begin{array}{c}(u-M\xi)x\\+(v-M\eta)y\end{array}\right]\right\}dxdy \quad (4)$$

where $M=-z_2/z_1$ and ignoring the constant term. The observation field can be calculated as follows.

$$U_i(u,v) = \mathcal{F}^{-1}\left[\mathcal{F}\left[\frac{1}{|M|}U_0\left(\frac{u}{M},\frac{v}{M}\right)\right]P(-\lambda z_2 f_x, -\lambda z_2 f_y)\right] \quad (5)$$

However, the approximation of the term associated with the object plane cannot be satisfied due to the vignetting effect in FPM. Thus the rigorous IRF is

$$h_c(u,v;\xi,\eta) = \exp\left[j\frac{k}{2z_2}(u^2+v^2)\right]\exp\left[j\frac{k}{2z_1}(\xi^2+\eta^2)\right]h_p(u,v) \quad (6)$$

It can be found that the $h_c$ is space-variant, which is corresponding to the ray-tracing method in Fig.1 (a). Thus the corresponding observation field can be rewritten as

$$U_i(u,v) = \exp\left[\frac{jk}{2z_2}(u^2+v^2)\right]\mathcal{F}^{-1}\left\{\mathcal{F}\left\{\frac{1}{|M|}U_0\left(\frac{u}{M},\frac{v}{M}\right)\exp\left[\frac{jk}{2z_1}\left(\frac{u^2+v^2}{M^2}\right)\right]\right\}P(-\lambda z_2 f_x, -\lambda z_2 f_y)\right\} \quad (7)$$

Figure 2 presents the raw images of angle-varied illumination in simulations and experiments. The experimental setup and data acquisition process for FPM can be found in the literature [1, 2] and will not be given unnecessary details here. A 15×15 programmable LED matrix with an illumination wavelength of 631.1nm, 20nm bandwidth and 4mm spacing is placed at 68.4mm above the sample stage. Generally, if the pupil function contains the center zero spot as shown in Fig.2 (e), the image is a bright field image, otherwise they are dark field images. Group (b) is the center 25 images in traditional LSI models with Eq.(5), while group (c) is those images captured by an 4×/0.1NA apochromatic objective and a 16-bits sCMOS (Neo 5.5, Andor, 6.5μm pixel pitch), which includes imperfect images and are different from group (b) due to the vignetting effect. In this letter, all the simulations and experiments use the same parameters. However, group (d) presents the results with our strict LSV model with Eq.(7), which can explain the phenomena of imperfect images. From an opposite perspective of ray trace, the LSV model modulates the spectrum of object in Eq.(7), enlarging it compared Fig.2 (e) with Fig.2 (f). When the pupil function translates with the angle-varied illumination, parts of the object spectrum will be cut if assuming the pupil function is fully illuminated and thus generating the imperfect images. This treatment of Eq.(7) will make the calculation more convenient.

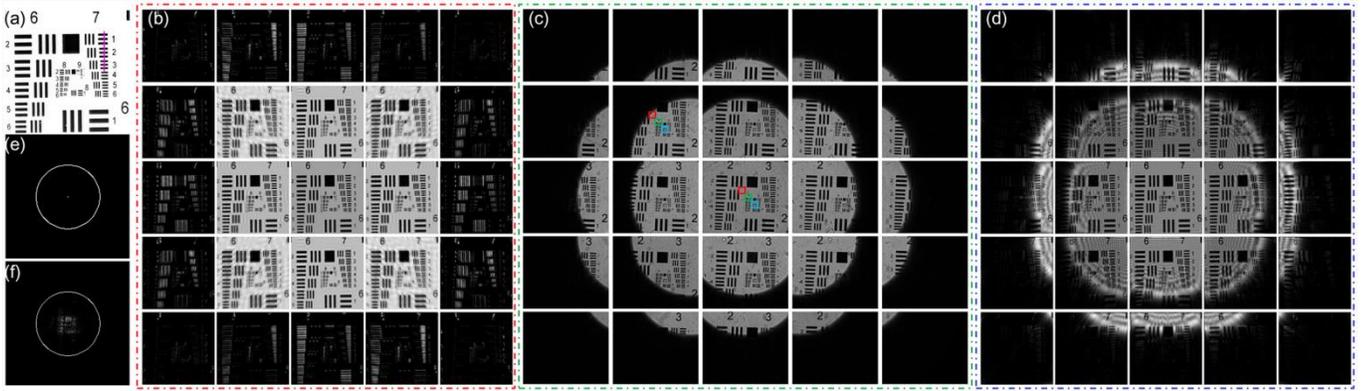

Fig. 2. The explanation of imperfect images with LSI and LSV models in simulations and experiments. (a) The USAF target as the ground truth for simulations. Group (b) 25 center images with LSI model in simulations. Group (c) 25 center images observed in experiments. Group (d) 25 center images with LSV model in simulations. (e) The spectrum of USAF target and pupil function (white circle) in LSI model without logarithmic operations. (f) The spectrum of USAF target and pupil function modulated by LSV model without logarithmic operations.

Figure 3 shows the profile of group 7, whose element 1 to 3 are in red, green and blue regions of Fig.2 (c), respectively. The half pitch resolution can reach at 3.16um in theory with 4×0.1NA objective [6, 18]. As shown in Fig.3 (a), with normal incidence, the resolution is invariable no matter which the regions are. Therefore, the effect of aberrations can be ignored for a power flat objective lens. And the Eq.(7) is equivalent to Eq.(5) in the non-vignetting area, which can also be deduced from the consistent results of the center image of Fig.2 (b) and Fig.2 (d). But in the vignetting area of Fig.2 (c) (top left of center image), the resolution is quite different as shown in Fig.3 (b), which gradually decreases from center to edge. Thus by comparing the Fig.3 (a) and Fig.3 (b), we can find that the resolution varies with space, which is not the reason for aberrations but for the vignetting effect.

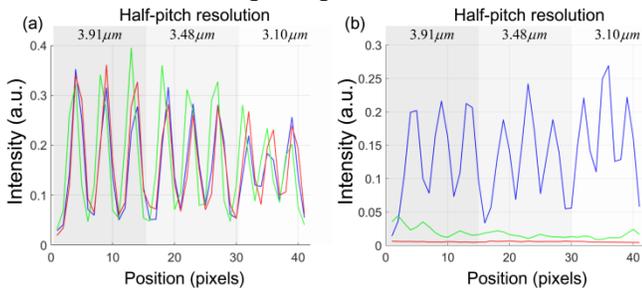

Fig. 3. The profiles of group 7, element 1 to 3 of red, green and blue region respectively of Fig.2.

Figure 4 (a) and (c) present the reconstructions of FPM in simulations and experiments, respectively, which will be deteriorated by the LSV model. Though the LSV model can explain the imperfect images, the model is also simplified. For example, there are apparent Gibb's phenomena [19] in Fig.2 (d) compared with the smooth edge in Fig.2 (c). In fact the objective usually consists of multiple optical elements; the inner structure of the optical path is complicated; the objective cannot be completely simulated with a single lens. Besides, some uncertain stray lights and the partial coherence of the LED light source cannot be evaluated precisely to the real conditions either. Thus it more proper to consider processing the LSV model approximate to LSI model by adding supplementary conditions. So the first strategy we propose is to divide the FOV into smaller segments, making the LSV model to local LSI model. And in fact, FPM really does the segmentation processing [1]. Although there are three benefits, parallel computing, smaller datasets and plane wave approximation, the reconstruction cannot be implemented without the segmentation processing as shown in Fig.4(c). Thus it also can be recognized that the model is space-variant indirectly. The smaller the segment is, the more possible the LSI approximation is, which explains why the segments in experiments will be no more than 400×400 pixels. However, there will be obvious wrinkles at the edge of FOV as shown in Fig.4 (a) and Fig.4 (d). It is because the imperfect images in the "transition zone" are not coincident with the LSI model, even being divided into small segments, which will also deteriorate the reconstruction quality severely. Here we propose the second strategy, omitting those imperfect images or adjusting the parameters to avoid appearing those imperfect images to bypass or partially alleviate the vignetting-induced reconstruction artifacts. Since comparing those BF images or DF images in both LSI with LSV models, we find that the differences are not huge in the small segments, but the imperfect images will have distant intensity differences. Though with the second strategy, there are still little differences between the BF images or DF images of LSI and LSV models, these differences will not over those aberrations, LED fluctuation, system parameter imperfection and noise [20-24]. And the EPRY-FPM [23] and adaptive step size strategy [24] will also play a compensatory or step-down role. In fact, many methods [25-27] only need 4 or 5 images to implement double resolution imaging. Thus omitting those imperfect images will not have any impact on the experimental results, and instead it makes the model meet the requirements of LSI as shown in Fig.4 (b) and Fig.4 (e). Figure 4 (b) presents the simulation results with our second strategy, which are a little blurry than the Fig.4(e), since there is no segment processing in the simulations. Comparing Fig.4 (a) with Fig.4 (b) and Fig.4 (d) with Fig.4 (e), we can find that the spectrum (red array) seems to be enlarged by the imperfect images. Otherwise if we use the LSV models, we must measure the coherent transfer function of each segment precisely. And the existing algorithms will be invalid due to the changes of objective functions and gradients, which will be complicated.

In addition, we also test our method in a biological sample (rabbit tongue section). Fig.5 (a) is the FOV. (a1) and (a2) are two segments with 100×100 and 200×200 pixels, respectively. By comparing (b1) and (c1), (b2) and (c2), we can conclude that our method can retrieve the phase at the edge very well and avoid appearing pleated artifacts.

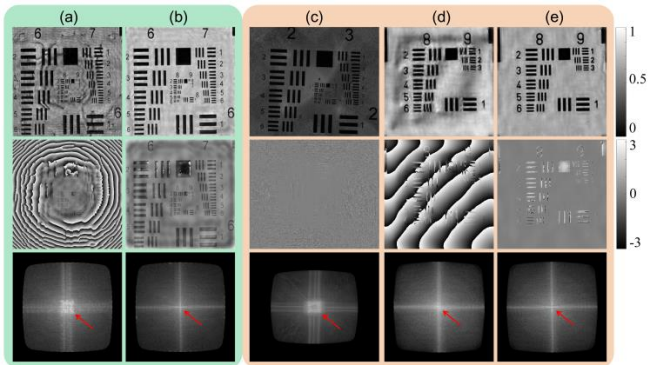

Fig. 4. The reconstruction of FPM at the edge. (a) Simulations without our method; (b) Simulations with our second strategy; (c) Experiments without our method; (d) One segment (50×50 pixels) in the center (c) with only segmentation processing; (e) Experiments with both of our strategies.

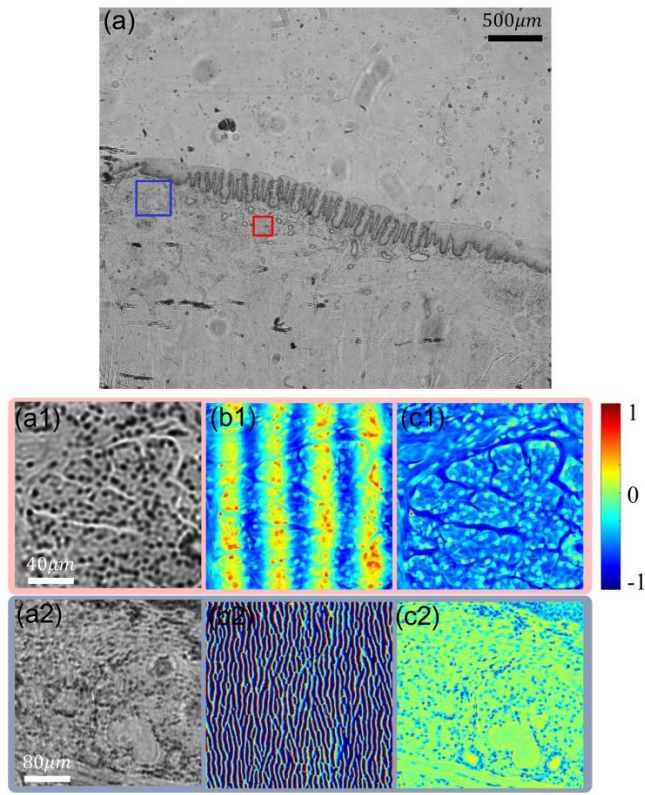

Fig. 5. The comparison of our method and original FPM with biological samples (rabbit tongue section). (a) The FOV captured by 4×0.1NA objective; (a1) and (a2) are two segments of (a); (b1) and (b2) present the phase results with original FPM, while (c1) and (c2) present the results with our method.

In conclusion, we examine the impact of this space-invariant approximation on FPM image formation from two different perspectives. Our analysis shows that for a practical FPM microscope with a low power objective and a large field of view, the space invariance is destroyed by diffraction at other stops associated with different lens elements to a large extent. A modified version of the LSV model is derived and discussed. By dividing the FOV into many small segments and omitting those imperfect images or adjusting the parameters, we can bypass or partially alleviate the vignetting-induced reconstruction artifacts. The effectiveness and performance of our methods are demonstrated in both simulations and experiments. The LSV models and strategies can be also conducive to other coherent imaging systems.

**Funding.** National Natural Science Foundation of China (NSFC) (61377008 and 81427802).

**Acknowledgment.** The authors thank Maosen Li for the valuable and helpful discussions and comments.